\documentclass[12pt]{article}
\usepackage{amssymb}
\usepackage{braket}
\usepackage{graphicx}

\begin{document}

\title{A Quantum-like Model of Selection Behavior}

\author{Masanari Asano \\
Liberal Arts Division, National Institute of Technology, Tokuyama College\\
 Gakuendai, Shunan, Yamaguchi 745-8585 Japan\\
Irina Basieva and Andrei Khrennikov\\
International Center for Mathematical Modeling\\
in Physics and Cognitive Sciences\\
Linnaeus University, V\"{a}xj\"{o}, Sweden\\
Masanori Ohya and  Yoshiharu Tanaka\\
Department of Information Sciences, Tokyo University of Science\\
Yamasaki 2641, Noda-shi, Chiba, 278-8510 Japan}

\maketitle

\begin{abstract}
In this paper, we introduce a new model of selection behavior under risk that describes an essential cognitive process for comparing values  of objects and making a selection decision. 
This model is constructed by the quantum-like approach that employs the state representation specific to quantum theory, which has the mathematical framework beyond the classical probability theory. 
We show that our quantum approach can clearly explain the famous examples of anomalies for the expected utility theory, the 
Ellsberg paradox, the Machina paradox and the disparity between WTA and WTP. Further, we point out that our model mathematically 
specifies the characteristics of the probability weighting function and the value function, which are 
basic concepts in the prospect theory.  
\end{abstract}

\section{Introduction}
\label{int}
Many studies on selection behavior have been done mainly in economics and psychology.
In economics, the \textit{expected utility theory} (\cite{VNG}) is traditionally used to discuss selection behaviors under risk, which are regarded as \textit{normative} 
and \textit{rational} from the view of the probability theory. In psychology, anomalies for the expected utility theory have been verified through a large number of experimental tests.\par
Among the two disciplines, behavioral economics based on the \textit{prospect theory} (\cite{KT,TK}) has been developed. The prospect theory is categorized into subjective expected 
utility (SEU) approach that tries to explain the anomalies by simulating the decision maker who makes a choice for maximizing the value of SEU. The SEU in the prospect theory is defined 
by the \textit{probability weighting function} and the \textit{value function}. The probability weighting function represents the psychological tendency to overestimate small 
probabilities and underestimate large ones. The value function represents the tendency that a loss gives a greater feeling of pain compared to the joy given by an equivalent gain. 
(The amount of loss or gain is measured from reference point whose position fluctuates situationally.) However, the development of experimental economics brought the finds of
 anomalies that cannot be captured in the prospect theory or its gentle modification. For examples, the anomalies shown by \cite{Ert},\cite{TJ}, \cite{Pay} and \cite{Bir} are well known and 
 they are difficult to explain. 
 
 In recent years, trying to find a theory/model that can explain all anomalies is a major topic in behavioral economics and, many researchers compete for developing descriptive decision-making 
model\footnote{Actually, Erev et al. \cite{Ert}  proposed and organized the fair competition of model (\textit{From Anomalies to Forecasts: Choice Prediction Competition for Decisions under 
Risk and Ambiguity} (CPC2015)). The detail of this competition is reported in the website http://departments.agri.huji.ac.il/cpc2015.}.\par
In this paper, we propose a new decision-making model that is not a gentle modification of SEU. It is a model designed in the \textit{quantum-like approach}, where 
the state representation specific to the \textit{quantum mechanics} is employed.\footnote{The problem of the state interpretation is one of the most complicated problems 
of quantum foundations. The present situation is characterized by a huge diversity of interpretations \cite{KHR5}. One of them is {\it the information interpretation} which was strongly supported by the 
recent quantum information revolution. We shall use this interpretation. Thus a quantum-like mental state represents information available to decision makers and  structured in 
the special  (quantum-like) way by their brains.}  Quantum mechanics is originally established for the description of statistical properties in microscopic phenomena.
The fundamental assertion in the quantum-like approach is that the formalism beyond the probability theory is also applicable to anomalies in various phenomena not limited to 
the microscopic cases.

We point out that there exist many quantum-like models designed under this assertion (\cite{AcS};  Accardi et al.\cite{Acc8,Acc9}; Asano et al.\cite{Asa11a,Asa11b,Asa11c,Asa12a,Asa12b}; \cite{Bas}; 
Busemeyer et al.\cite{Bus1,Bus2,Bus3,Bus4,Bus5}; Cheon \& Takahashi\cite{CHT1,CHT2}; Conte et al.\cite{CO1,CO2,CO3}; \cite{DK}; Haven \& Khrennikov\cite{HK1,HK2}; 
Khrennikov et al.\cite{KHR1,KHR2,KHR3,KHR4,KHR5,KHR6,KHRR,KHR7,KHR8,KHR9}; \cite{OV}; \cite{PB}).\footnote{The quantum-like approach works very well to model a variety 
of behavioral effects, e.g., the order effect or disjunction and conjunction effects. However, it is not clear whether the mathematical formalism of quantum theory can serve to model 
all possible behavioral phenomena, see Khrennikov et al.  \cite{KHR9} 
and Boyer-Kassem et al. \cite{Boyer1}, \cite{Boyer2} for the recent analysis of this problem.} 

Our main interest is not just to make a model that fits experimental data of anomalies, but {\it to offer the foundation for the new theory of expected theory,} 
that is, to describe human behavior using non-classical probabilities. 
We believe that in order to develop the quantum-like approach further to become an established theory,  we need to investigate in a deep way how the quantum-like approach compares 
to prospect theory.  In this paper, we will show that the characters of the probability weighting function and the value function can be realized mathematically in a part of our quantum-like model. 
Further, we will show that our model can explain several anomalies including non-classical ones. These results suggest that our quantum-like model has the potential to be a mile stone toward the development of model to cover all known and yet unknown anomalies.\par
In Sec.~2, we mathematically define a cognitive process essential to make a preference.
Firstly, the decision maker in our model is aware of the existence of objects to be compared.
The structure of awareness is represented by \textit{density operator(matrix)}, which is the most general state representation in quantum mechanics. We call it the \textit{comparison state}.
The decision maker secondly compares the values (utilities) of objects quantitatively.
This functionality is represented as a mapping from the comparison state to a real number, which is called the \textit{evaluation function}. A selection decision is derived from the above comparison state and evaluation function.  
In Sec.~2.3, we discuss the modeling of selection between two lotteries. The main problem is how to embed the probability distributions into the comparison state. Especially, the comparison state designed in Sec.~2.3.3 is important, because it is closely related with the crucial concept 
in quantum theory: The \textit{state transition}, which occurs when a physical value is measured on a system, is the basic assumption in quantum theory. A physical state before measurement is generally represented with the form called \textit{quantum superposition} that is clearly distinguished from a statistical description as obtained after measurement. If the comparison state at the stage before drawing the lots
 is represented with quantum superposition, the character of the probability weighting function can be explained, see Fig.~1 in Sec.~2.3.4.\par 
It is also crucial that our model describes the cognitive process that is never explained in the expected utility theory. For example, for the lottery that pays $\$100$ or $\$1$ according to a probability distribution, a decision maker might fear to get $\$1$ by missing the chance of $\$100$. Then, the difference of these potential outcomes will 
affect his/her \textit{evaluation of risk}. Such a process is to be experienced before drawing the lottery but never after drawing, and actually, its effect is represented in the comparison state with quantum superposition, as the parameter called \textit{degree of evaluation of risk}(DER).
We expect, the evaluation of risk is an essential cause of anomaly. In Sec.3, this point will be shown clearly, where we simulate the famous anomalies in 
selection behavior under \textit{ambiguity}; \textit{Ellsberg paradox}(\cite{EL}) 
and the \textit{Machina paradox}(\cite{MC})\footnote{There exists another quantum-like approach trying to solve the two paradoxes, see (\cite{AS}), in which the effect of DER is not assumed.}.
Ellsberg paradox, see Table.~1 is the first example that shows the anomaly due to the \textit{ambiguity aversion}, 
which lets one to prefer the known risk to the unknown risk.  The Machina paradox, see Table.~3 points out an existence of anomaly that is impossible to be
explained even in the popular models of ambiguity aversion. Our analyses are summarized in the diagrams of Fig.~2 and Fig.~3 in Sec.~3.\par 
In Sec.~4, we discuss the determination of \textit{cash equivalent} (CE), which is an amount of money whose value is indifferent from a given lot. 
The CE is related to the \textit{willingness to accept} (WTA) and \textit{willingness to pay} (WTP) (\cite{HR}), each of which is interpreted as cash determined in the aim of selling or buying the lot.
It is well-known that the relation of WTA$>$WTP is generally observed in experimental tests, and this disparity is an important topic in economics. As seen in Fig.~4, CE is defined  as the function of the degree of evaluation of risk (DER). Therefore, we can explain the disparity from DER whose value is to be changed 
 depending  on the decision maker's situation. Note that in such a situation dependency is consistent with the character of value function in the prospect theory.

\section{A Model of Selection Behavior}  
\textit{There are two lots, say $A$ and $B$. If you chose $A$, you will get outcome $x_i$ ($i=1\cdots n$) with probability $P_i$. If you chose $B$, you will get outcome $x_i$ with probability $Q_i$. 
All of the outcomes are different from each other. Which lot do you select?}\\
\\
When a decision maker decides the preference $A\succ B$ or $B \succ A$ in this situation, he/she will recognize the following three points. \\ 
1. \textit{What} objects exist,\\
2. \textit{Which} pairs of objects are to be compared,\\
3. \textit{How} comparisons are evaluated.\\
Here, we have to emphasize that ``objects" mean ``events" that will be experienced in the future. He/she can simulate the experience that he draws the lot $ A $ ($ B $) and get the outcome $ x_i $. Let us 
represent such an event by $ (A,x_i) $ or $(B,x_i) $. Further, we assume that the decision maker sets the utilities of $ (A,x_i) $ and $(B,x_i) $ by $ u(x_i)\equiv u_i $. (The utility of event depends 
on only outcome.) Here, $ u(x) $ is a utility function of outcome $ x $. Under this assumption, comparing an arbitrary 
pair of objects (events) becomes possible. Lastly, the decision maker evaluates 
various comparisons for making the preference $A\succ B$ or $B \succ A$.
If von Neumann-Morgenstern (VNM) utility theorem is applied to this third point, the method consistent with VNM axioms (Completeness, Transitivity, Independence, Continuity) will be used. 
VNM axioms are given for the relation of utilities like $ u\succ v $ and the operation using probability like $pu+(1-p)v$. Therefore, these encourage the decision maker to estimate the 
expected utilities, $E_A=\sum_i^n P_i u_i $ and $E_B=\sum_i^n Q_i u_i$, and employ its difference as 
a sort of a criterion for making the preference. \par 
However, the fact of anomalies shows that many people seem to violate VNM axioms in their selection behaviors. For this problem, we have to mention the interpretation of probability used in the axioms. In the book ``Theory of Games and Economic Behavior"(\cite{VNG}), VNM stressed the well founded interpretation of
 probability as \textit{frequency}, but not a \textit{subjective concept of probability}, which is often visualized. 

{\it What is the subjective concept?} It is not the frequency of event counted through a large number of trial experiments, rather it is the \textit{weight of awareness} assigned for unmeasured event. 
The reason that many people often visualize the latter impression might be simple: It is difficult for them to simulate a large number of trial experiments in their minds, 
even if they know the meaning of frequency probability. We believe, in a realistic selection behavior using 
the subjective probability, a ``natural" operation exists, 
which is different from the form of $pu+(1-p)v$, and in this section, we describe it by using the framework of quantum theory. It is a significant point that quantum theory 
mathematically defines a state before measurement. Here, one can interpret the state as a subjective recognition for uncertain (unmeasured) events, if the measurement is regarded as an 
acquisition of subjective experience. Such an interpretation of quantum theory is called ``Qbism", see the papers of \cite{FS}. (Note, QBism never deny the standard interpretation 
in quantum physics, rather, includes it.) In Sec.2.1 and Sec.2.2, the three points introduced in the above are explained in the state 
representation based on QBism, and in Sec.2.3, a criterion for selection is defined as a ``natural" operation with potential to violate VNM axioms.

\subsection{Comparison State}
Firstly, the decision maker is aware of existences of objects. Secondly, he/she is aware of various pairs of objects to be compared. As mentioned in the beginning of section, the ``objects" we consider here are the ``events" individual and incompatible each other. The decision maker preliminarily assigns his/her awarenesses for these events. In our model, its distribution is mathematically 
represented in the form of a \textit{density operator}, which is the general state
 representation in quantum theory. In a sense, this representation plays the role of\textit{ prior probability distribution} that is different from statistical probability distribution.\par 
The density operator $\rho$ we use here is defined as a $N\times N$ complex matrix on a finite dimensional complex vector space $\mathcal{H}=\mathbb{C}^N$, which is described as
\begin{equation}
\rho=\sum_{k,l}^N \rho_{kl} |k \rangle \langle l| =\sum_{k,l}^N \rho_{kl} t_{kl}.\label{dens} 
\end{equation}
The notation $\ket{k}$ ($\bra{k}$), which is called ket vector (bra vector), means a column vector (row vector) with the $k$-th component is $1,$ and other components are zero.
Thus, $\{\ket{k}\}$($k=1\cdots N$) is a set of vectors that are normalized and orthogonal to each other; $\braket{k|l}=\delta_{kl}$. ($\braket{k|l}$ means the inner product of two vectors.) 
Hereafter, we call the set $\{\ket{k}\}$ a \textit{complete orthonormal system} (CONS) in $\mathcal{H}$.
$\braket{k|\rho|l}=\rho_{kl}\in \mathbb{C}$ is $(k,l)$-component of the matrix $\rho$. The notation $t_{kl}$  denotes the operator $|k\rangle \langle l|$, 
which acts as follows: $t_{kl}\ket{h}=\delta_{lh}\ket{k}$. In general, a density operator satisfies the following conditions:
\begin{enumerate}
  \item $\rho_{kl}=\rho_{lk}^*$ ; the self-adjoint is satisfied. 
  \item $tr(\rho)=1$ ; the diagonal sum is one.
  \item $\forall x\in \mathcal{H}$, $\left<x\right| \rho \left|x\right>\ge 0$ ; the positivity is satisfied.
\end{enumerate}
($\rho_{lk}^*$ means the complex conjugate of $\rho_{lk}$.)\par
We divide this matrix into the diagonal part and non-diagonal part;
\begin{equation}
\rho=\rho_{d}+\rho_{nd}
\end{equation}
where
\[\rho_{d}=\sum_k^N \rho_{kk}t_{kk}=\sum_k^N \rho_{kk}E_k\]
\[\rho_{nd}=\sum_{k<l}\rho_{kl}t_{kl}+\rho_{lk}t_{lk}=\sum_{k<l}|\rho_{kl}|(\rm{e}^{i \theta_{\it{kl}}}\it{t_{kl}}+\rm{e}^{- \rm{i}\theta_{\it{kl}}}\it{t_{lk}})=\sum_{k<l}|\rho_{kl}|C_{k\leftrightarrow l}\]
From the above three conditions, the components $\{\rho_{kk} \}$ in $\rho_{d}$ are real numbers with $\rho_{kk}>0$ and $\sum_k^N \rho_{kk}=1$. The components $\rho_{kl}$ in $\rho_{nd}$ are generally complex numbers, $\rho_{kl}=|\rho_{kl}|\rm{e}^{i\theta_{\it{kl}}}$. The operators $t_{kk}=E_k$ and $\rm{e}^{\rm{i}\theta_{\it{kl}}}\it{t_{kl}}+\rm{e}^{- \rm{i}\theta_{\it{kl}}}\it{t_{lk}}=C_{k\leftrightarrow l}$ in the above equations are also called \textit{transition operator}s: $E_k$ transits the vector $\ket{k}$ to itself:
\[E_k\ket{h}=|k\rangle \langle k| h \rangle =\delta_{kh}\ket{k}. \]
$C_{k\leftrightarrow l}$ transits $\ket{k}$ to $\rm{e}^{-\rm{i}\theta_{\it{kl}} }\ket{\it{l}}$ or $\ket{l}$ to $\rm{e}^{\rm{i}\theta_{\it{kl}} } \ket{\textit{k}}$: 
\[C_{k\leftrightarrow l}\ket{h}=\delta_{lh}\rm{e}^{\rm{i}\theta_{\it{kl}} } \ket{\textit{k}}+\delta_{\it{kh}}\rm{e}^{-\rm{i}\theta_{\it{kl}} }\ket{\it{l}}.\]
We interpret the structure of $\rho$ as follows: The CONS $\{\ket{k}\}$ corresponds to $N$ objects. The transition operator $E_k$ 
represents awareness of existence of object-$k$ and $C_{k\leftrightarrow l}$ represents
 awareness of a pair of objects $k$ and $l$. The values of $\rho_{kk}$ and $|\rho_{kl} |$ 
imply the ``weights" of awareness assigned for these cognitive processes. We call the density matrix $\rho$ the \textit{comparison state}, hereafter. 

\subsubsection*{Simple Example of Comparison State}
Let us introduce a most simple comparison state with $N=2$. The decision maker is aware of two objects and recognizes them as one pair to be compared. Then, the following $2\times 2$ matrix on $\mathcal{H}=\mathbb{C}^2$ is considered as a form of comparison state.
\begin{equation}
\rho=\left(
  \begin{array}{cc}
    \alpha^2   & \alpha \beta \rm{e}^{i\theta_{12}}   \\
    \alpha \beta \rm{e}^{-i\theta_{12}}   & \beta^2   \\
  \end{array}
\right)
=(\alpha^2 E_1+ \beta^2 E_2)+ \alpha \beta C_{1\leftrightarrow 2}\label{CS1}
\end{equation}
$\alpha$ and $\beta$ are real numbers satisfying $\alpha^2+\beta^2=1$. The weights $\alpha^2$, $\beta^2$ and $\alpha \beta$ are assigned for the cognitive processes $E_{1,2}$ and  $C_{1\leftrightarrow 2}$\footnote{The form of Eq.~(\ref{CS1}) is rewritten as
\[
\rho=|\psi \rangle \langle \psi|,\ \ket{\psi}=\alpha \rm{e}^{i\theta_{12}} \ket{1}+\beta \ket{2}=\left(
  \begin{array}{c}
    \alpha \rm{e}^{i\theta_{12}}  \\
    \beta   \\
  \end{array}
\right).
\]
Such  representation, which is called \textit{pure state} in quantum theory, is used in the general construction of comparison state, see Sec.~2.3.
}. 
In principle, the values of $\alpha$ and $\beta$ are to be determined subjectively by the decision maker. In the case of $(\alpha,\beta)=(0,1)$ or $(1,0)$, $\alpha \beta$ is zero. This 
implies that the decision maker recognizes the pair of objects, if and only if he/she is aware of the existence 
of  both objects. Also, the condition of $(\alpha,\beta)=(\frac{1}{\sqrt{2}},\frac{1}{\sqrt{2}})$ implies 
that the decision maker pays attention to the existence of them,  in a \textit{fair way}.\par

\subsection{Evaluation Function}
The decision maker we postulate has a function to compare an arbitrarily pair which is chosen from $N$ objects. It is the quantitative evaluation defined by the following map.
\begin{equation}
\varphi_D:\mathfrak{S}(\mathcal{H}) \mapsto \mathbb{R},\ \varphi_D(\cdot):=tr(D\cdot).\label{MEF}
\end{equation}
$\mathfrak{S}(\mathcal{H})$ denotes the set of all of comparison states (density matrices) on $\mathcal{H}=\mathbb{C}^N$.
The operator $D$ is defined by a self-adjoint operator \footnote{In the formalism of quantum theory, a measurable 
physical quantity is given by a self-adjoint operator like $D$, and the value of $tr(D\rho)$ is defined 
as the expectation value of $D$ for the state $\rho$.}, which is also written as a $N\times N$ complex matrix; 
\begin{equation}
D=\sum_{k,l}^N F(k,l)|k\rangle \langle l|=\sum_{k,l}^N F(k,l)t_{kl}
\end{equation}
$F(k,l) \in \mathbb{C}$ is $(k,l)$-component of the matrix. From the self-adjoint property, $F(k,l)^*=F(l,k)$ is satisfied. The objects $ k $ and $ l $ are events independent and incompatible each other. (The orthogonality $ \langle k|l\rangle =\delta_{k,l} $ implies this fact.)  However, the decision maker correlates them in his/her cognitive operation of comparison. 
The correlation, that is, how the decision maker sees the difference between $k$ and $l$ is encoded 
in the term of $ F(k,l) = \langle k|D|l\rangle $. We give its form as
\begin{eqnarray}
F(k,l)=(u_k-u_l)\rm{e}^{i\phi_{\it{kl}}},\label{EF}
\end{eqnarray}
where $u_k$ is the utility that the decision maker feels for the object-$k$.
Hereafter, we call $\varphi_D$ the \textit{evaluation function}.

\subsubsection*{Simple Example of Evaluation}
Let us explain the use of evaluation function $\varphi_D(\rho)$ in the selection between two objects: The comparison state $\rho$ is given in the form of Eq.~(\ref{CS1}):
\[\rho=\frac{1}{2}\left(
  \begin{array}{cc}
    1   &  \rm{e}^{i\theta_{12}}  \\
     \rm{e}^{-i\theta_{12}}  &  1  \\
  \end{array}
\right).
\]
where the fairness $(\alpha, \beta)=(\frac{1}{\sqrt{2}},\frac{1}{\sqrt{2}})$ is assumed. From the definition of Eq.~(\ref{EF}), the operator $D$ is given as $2\times 2$ matrix as
\[D=\left(
  \begin{array}{cc}
    0   &  (u_1-u_2)\rm{e}^{i\phi_{12}}  \\
    (u_1-u_2)\rm{e}^{-i\phi_{12}}   &  0  \\
  \end{array}
\right).
\]
From the calculation
\[D\rho=\frac{u_1-u_2}{2}\left(
  \begin{array}{cc}
    \rm{e}^{i(\phi_{12}-\theta_{12})}   &\rm{e}^{i\phi_{12}}   \\
     \rm{e}^{-i\phi_{12}}  & \rm{e}^{-i(\phi_{12}-\theta_{12})}  \\
  \end{array}
\right),
\] 
the evaluation 
\[\varphi_D(\rho)=tr(D\rho)=\cos \Theta_{12} (u_1-u_2),\ \Theta_{12}=\phi_{12}-\theta_{12}\]
is obtained.
The parameter $\Theta_{12}$ determines the \textit{direction for evaluation}. For example, in the case of $\cos \Theta_{12}>0$, the above $\varphi_D(\rho)$ is used in the following way.
\[If\ \varphi_D(\rho)\ge 0,\ then \ 1\succeq 2,\ If\ \varphi_D(\rho)\le 0,\ then \ 2\succeq 1. \]
In the case of $\cos \Theta_{12}<0$,
\[If\ \varphi_D(\rho)\le 0,\ then \ 1\succeq 2,\ If\ \varphi_D(\rho)\ge 0,\ then \ 2\succeq 1. \]

\subsection{Criterion for Selection}
 Let us remind the situation in the beginning of section: If a decision maker draws the lot $A$ ($B$), he/she will get the outcome $x_i$ ($ i=1,\cdots ,n $) with the probability $P_i$ ($Q_i$). We defined the objects in this situation as the events (experiences),
\[ \set{(A,x_1),\cdots,(A,x_n),(B,x_1),\cdots , (B,x_n)}. \]
(For example, $ (A,x_k) $ denotes the event (experience) that he/she draws the lot $ A $ and gets the outcome $ x_k $.) In our modeling, these are to be represented by vectors in
 Hilbert space. Firstly, 
let us introduce the \textit{tensor product of spaces}, $\mathcal{H}=\mathcal{H}_{Lot}\otimes \mathcal{H}_{Outcome}$; $\mathcal{H}_{Lot}$ is the two dimensional space spanned by the two vectors $\ket{A}$ and $\ket{B}$, and $\mathcal{H}_{Outcome}$ is the $ n $ dimensional space spanned by $\{\ket{x_i},i=1,\cdots,n\}$. Here, $\braket{A|B}=\braket{B|A}=0,\ \braket{x_i|x_j}=\delta_{i,j}$ are assumed, that is, 
\[\{\ket{A}\otimes \ket{x_1},\cdots,\ket{A}\otimes \ket{x_n} ,\ket{B}\otimes \ket{x_1},\cdots,\ket{B}\otimes \ket{x_n}\}.\] 
 is a CONS of $\mathcal{H}=\mathbb{C}^{2n}$. These vectors correspond to the above objects.
Hereafter, $\{\ket{A}\otimes \ket{x_i}\}$ and $\{\ket{B}\otimes \ket{x_i}\}$
are replaced by $\{\ket{i},i=1,\cdots , n\}$ and $\{\ket{\bar{i}}=\ket{n+i},i=1,\cdots , n\}$: We call the events (experiences) $(A,x_i)$ and $(B,x_i)$, object-$i$ and object-$\bar{i}$. (The utility of object-$ i $ or object-$ \bar{i} $ is given by $ u(x_i)=u_i $.)\par  
The comparison state $\rho$ can be defined in $\mathcal{H}=\mathbb{C}^N$ with $ N=2n$. Then, the statistical information of $\set{P_i}$ and $\set{Q_i}$ will be embedded in its structure. In this subsection, the two types of $\rho$ are designed under different interpretations,
 and in each case, the evaluation $\varphi_D(\rho)$ is used as the criterion for selection between $A$ and $B$. 
 
\subsubsection{Classical Comparison State}
\textit{After you draw the two lots, you will recognize the experiences $(A,x_i)$ and $(B,x_j)$ with the probability $ P_i\times Q_j$.}\\ 
This context is explained in the following comparison state.
\begin{eqnarray}
\rho_C=\sum_{i,j=1}^{n} P_i Q_j |\psi_{ij} \rangle \langle \psi_{ij}|\in \mathfrak{S}(\mathcal{H}), \label{CS}
\end{eqnarray}
where $\ket{\psi_{ij}}$ is the normalized vector with the form of 
\[\ket{\psi_{ij}}=\frac{1}{\sqrt{2}}(\ket{A}\otimes \ket{x_i}+\ket{B}\otimes \ket{x_j})=\frac{1}{\sqrt{2}}(\ket{i}+\ket{\bar{j}}),\]
where $ \bar{j}=n+j $. The state $\rho_C$ has the form of so-called \textit{mixed state}\footnote{Generally, a mixed state is defined as $\rho=\sum_k^M \alpha_k |\psi_k \rangle \langle \psi_k|$, where $\set{\alpha_k}$ is a probability distribution satisfying $\sum_k^M \alpha_k=1$, and $\set{\ket{\psi_k}\in \mathbb{C}^N}$ are $N$-dimensional vectors whose norms are one.}.  
Note that $|\psi_{ij}\rangle \langle\psi_{ij}|$ in $\rho_C$ is rewritten as
\[|\psi_{ij}\rangle \langle\psi_{ij}|=\frac{1}{2}(E_i+E_{\bar{j}})+\frac{1}{2}C_{i\leftrightarrow \bar{j}},\]
which is the same with the state of Eq.~(\ref{CS1}) with $(\alpha,\beta)=(\frac{1}{\sqrt{2}},\frac{1}{\sqrt{2}})$:
 The decision maker at $|\psi_{ij}\rangle \langle\psi_{ij}|$ is aware of the existence 
of objects denoted by $i$, $\bar{j}$ and recognizes them as one pair. 
This is interpreted as the cognitive state that is experienced \textit{after} drawing the two lots. 

\subsubsection{Classical Criterion}
The evaluation of the comparison state $\rho_C$ is calculated as
\begin{eqnarray}
\varphi_D(\rho_C)=\sum_{i,j=1}^{n}P_iQ_j\varphi_D(|\psi_{ij}\rangle \langle\psi_{ij}|)=\sum_{i,j=1}^{n}P_iQ_j\cos \Theta_{i\bar{j}}(u_i-u_{j}).\label{CCSE}
\end{eqnarray}
Here, we used the operator $D$ of Eq.~(\ref{EF}). 
The value of $\varphi_D(\rho_C)$ is the statistical expectation of $\set{\varphi_D(|\psi_{ij}\rangle \langle\psi_{ij}|)}$.
Especially, if all of $\set{\Theta_{i\bar{j}}}$ have the same value $\Theta$, it is proportional to the difference of expected utilities;
\begin{equation}
\varphi_D(\rho_C)=\cos \Theta(\sum_{i=1}^{n}P_i u_i -\sum_{j=1}^{n}Q_j u_j)\propto E_A-E_B.\label{EP}
\end{equation}
As result, our model can realize the criterion based on the expected utility theory as one example.
\subsubsection{Non-classical Comparison State}
The classical comparison state $\rho_C$ of Eq.~(\ref{CS}) has the form as the statistical description of the cognitive states $\{|\psi_{ij}\rangle \langle\psi_{ij}|\}$, which are experienced after drawing the lots. 
In this subsection, we discuss another comparison state that describes cognitive processes \textit{before drawing}. 
We start this discussion by regarding the action of drawing of lot as a sort of \textit{measurement}. 
Acquisition of information through a measurement resolves \textit{uncertainty}, which an observer holds before measurement. 
In other word, by the measurement, awareness of uncertainty is transited to the one of definite information.  
We believe that such transition is described in the mathematical formalism of quantum theory, which
have discussed state changes by measurement, generally.
 In our description,  awareness of uncertainties on the two lots is represented by the following two density operators.
\begin{eqnarray} 
\sigma_A=|\psi_A \rangle \langle \psi_A|,\ \sigma_B=|\psi_B \rangle \langle \psi_B|,
\end{eqnarray}
where $\ket{\psi_A}$ and $\ket{\psi_B}$ are defined as normalized vectors in $\mathcal{H}=\mathbb{C}^{N=2n}$.
If the decision maker draws the lot $A$ and knows the result of event $(A,x_i)$ (object-$i$), for example, the uncertainty symbolized by $\sigma_A$ is 
transited to the definite cognition $|i \rangle \langle i|$. According to
 quantum theory, the state transition from $\sigma_A$ to $|i \rangle \langle i|$ is represented by
\[\frac{M_i \sigma_A M_i^*}{tr(M_i \sigma_A M_i^*)}=|i \rangle \langle i|,\]
where $M_i(i=1,\cdots ,N)$ is called measurement operator and defined by $M_i=|i \rangle \langle i|$. 
Further, the statistical description obtained after measurements $\set{M_i}$ is defined by
\[\sum_{i=1}^{n} M_i \sigma_A M_i^*=\sum_{i=1}^{n} tr(M_i \sigma_A M_i^*) |i \rangle \langle i|=\sum_{i=1}^{n} |\braket{i|\psi_A}|^2 |i \rangle \langle i|. \]
Here, the value of $tr(M_i \sigma_A M_i^*)$ corresponds to the probability of measurement of result $i$:
\[tr(M_i \sigma_A M_i^*)=|\braket{i|\psi_A}|^2=P_i.\] 
Thus, the vector $\ket{\psi_A}$ in $\sigma_A$ is written as
\begin{equation}
\ket{\psi_A}=\sum_{i=1}^{n} \sqrt{P_i} \rm{e}^{i\theta_{\it{i}}^A}\it{\ket{i}}.\label{V1}
\end{equation}
This form is called \textit{quantum superposition}.
Similarly, $\ket{\psi_{B}}$ in $\sigma_B$ has the form of
\begin{equation}
\ket{\psi_B}=\sum_{i=1}^{n} \sqrt{Q_i} \rm{e}^{i\theta_{\bar{\it{i}}}^B}\it{\ket{\bar{i}}}.\label{V2}
\end{equation}
As seen in these results, the probability distributions $\set{P_i}$ and $\set{Q_j}$ are embedded as elements determining the directions of $\ket{\psi_{A,B}}$.\par
With using the above $\ket{\psi_{A,B}}$, we define the following comparison state: 
\begin{equation}
\rho=|\Psi \rangle \langle \Psi|,\ \ket{\Psi}=\frac{1}{\sqrt{2}}(\ket{\psi_{A}}+\ket{\psi_{B}})\label{QLC0}.\nonumber
\end{equation}
Note, this $\rho$ is rewritten as
\begin{eqnarray}
\rho=\frac{1}{2}(\sigma_A+\sigma_B)+\frac{1}{2}C_{A\leftrightarrow B}.\label{QLC}
\end{eqnarray}
where
\[C_{A\leftrightarrow B}=|\psi_A\rangle \langle \psi_B|+|\psi_B\rangle \langle \psi_A|.\]
The form of Eq~(\ref{QLC}) clearly explains the basic cognitive processes for the selection between $A$ and $B$ at the 
stage before measurement: the term of $\frac{1}{2}(\sigma_A+\sigma_B)$ in Eq.~(\ref{QLC}) implies that the 
decision maker is aware of the existence of two lots. The term of $C_{A\leftrightarrow B}$ implies that he/she recognizes the 
two lots as one pair to be compared.  

\subsubsection{Non-classical Criterion}
Let us discuss the evaluation of the non-classical comparison state $\rho$;
\begin{eqnarray}
\varphi_D(\rho)=\varphi_D \left(\frac{1}{2}\sigma_A \right)+\varphi_D \left(\frac{1}{2}\sigma_B \right)+\varphi_D \left(\frac{1}{2}C_{A\leftrightarrow B} \right).\label{ENC}
\end{eqnarray}
Here, we use the evaluation function $\varphi_D$ with $D$ of Eq.~(\ref{EF}).
The first and second terms $\varphi_D \left(\frac{1}{2}\sigma_A \right)$ and $\varphi_D \left(\frac{1}{2}\sigma_B \right)$ are calculated as
\begin{eqnarray}
\varphi_D \left(\frac{1}{2}\sigma_A \right)&=&\sum_{i, j=1}^{n}\sqrt{P_iP_{j}}\cos \Theta_{ij} (u_i-u_{j})\nonumber \\
\varphi_D \left(\frac{1}{2}\sigma_B \right)&=&\sum_{i, j=1}^{n}\sqrt{Q_iQ_j}\cos \Theta_{\bar{i}\bar{j}} (u_i-u_j),\label{ERisk}
\end{eqnarray}
where $\Theta_{ij}=\phi_{ij}-\theta_{i}^A+\theta_{j}^A$ and $\Theta_{\bar{i}\bar{j}}=\phi_{\bar{i}\bar{j}}-\theta_{\bar{i}}^B+\theta_{\bar{j}}^B$.
We call these terms the \textit{evaluations of risks}.
For example, if the outcome in the lot A is $\$ 100$ with probability $P$ and $\$10$ with probability $1-P$,
a decision maker might feel the risk by seeing the difference between $\$ 100$ and $\$10$, and he/she will estimate the highest risk when $P=1/2$. 
In this way, a decision maker will evaluate the risk from the differences of outcomes and probabilities. As seen in the form of Eq.~(\ref{ERisk}), the degrees of evaluations of risks are specified by the values of $\set{\cos \Theta_{ij}}$ and $\set{\cos \Theta_{\bar{i}\bar{j}}}$. Hereafter, we call them DERs.  
The evaluation of risk seems to be experienced in realistic selection behavior.
However, in the classical criterion $\varphi_D(\rho_C)$, this empirical fact is ignored, see Sec.~2.3.2.\par
The third term $\varphi_D \left(\frac{1}{2}C_{A\leftrightarrow B}\right)$ in Eq.~(\ref{ENC}) is calculated as
\begin{eqnarray}
\varphi_D \left(\frac{1}{2}C_{A\leftrightarrow B}\right)&=&\sum_{i, j=1}^{n}\sqrt{P_iQ_{j}}\cos \Theta_{i\bar{j}} (u_i-u_{j}),\label{ECom}
\end{eqnarray}
where $\Theta_{i\bar{j}}=\phi_{i\bar{j}}-\theta_{i}^A+\theta_{\bar{j}}^B$. 
This term corresponds to the classical criterion $\varphi_D(\rho_C)=\sum_{i,j}P_iQ_j\cos \Theta_{i\bar{j}}(u_i-u_{j})$ of Eq.~(\ref{CCSE}). 
It should be noted here that the weights of awareness in Eq.~(\ref{ECom}) are given by square roots of products of probabilities.
In the case of $\Theta_{i\bar{j}}=\Theta$, the classical criterion realizes the selection based on the expected utility theory, see Eq.~(\ref{EP}).
In the same case, Eq.~(\ref{ECom}) is written as
\begin{eqnarray}
\varphi_D \left(\frac{1}{2}C_{A\leftrightarrow B}\right)=\cos \Theta \left[ \sum_{i, j=1}^{n}\sqrt{P_iQ_{j}} u_i- \sum_{i, j=1}^{n}\sqrt{P_iQ_{j}} u_j \right].\nonumber
\end{eqnarray}
Here, we consider the following value:
\begin{eqnarray}
\frac{\varphi_D \left(\frac{1}{2}C_{A\leftrightarrow B}\right)}{\sum_{k ,l}\sqrt{P_k Q_l}}&=& \cos\Theta \left[\sum_{i=1}^{n}\frac{\sqrt{P_i}}{\sum_{k}\sqrt{P_k}} u_i- \sum_{ j=1}^{n}\frac{\sqrt{Q_{j}}}{\sum_{l}Q_{l}} u_j\right] \nonumber \\
&=&\cos\Theta \left[\sum_{i=1}^{n}\tilde{P}_i u_i- \sum_{ j=1}^{n}\tilde{Q}_j u_j\right].
\end{eqnarray}
Note, $\sum_{i=1}^n \tilde{P}_i=1$ and $\sum_{j=1}^{n}\tilde{Q}_j=1$. From the above equation, one can see that the evaluation $\varphi_{D}\left(\frac{1}{2}C_{A\leftrightarrow B}\right)$ is essentially equivalent to the calculation of expected utilities, $\sum_{i}\tilde{P}_i u_i$ and $\sum_{ j}\tilde{Q}_j u_j$. 
We call $\set{\tilde{P}_i}$ and $\set{\tilde{Q}_i}$, which are different from the statistical probabilities, \textit{subjective probabilities}.
The subjective probability $\tilde{P}_i$ is rewritten as 
\begin{eqnarray}
\tilde{P}_i=\frac{\sqrt{P_i}}{\sqrt{P_i}+\left(\sum_{k\neq i}\sqrt{\frac{P_k}{1-P_i}}\right)\sqrt{1-P_i}}=\frac{\sqrt{P_i}}{\sqrt{P_i}+\xi\sqrt{1-P_i}}.
\end{eqnarray}
The values of $\set{\frac{P_{k\neq i}}{1-P_i}}$ in the above equation mean the probability distribution of results except for $i$. The range of $\xi=\sum_{k\neq i}\sqrt{\frac{P_k}{1-P_i}}$ is $1\le \xi \le \sqrt{m-1}$. 
Let us explain the relation between statistical and subjective probabilities by using the function,
\begin{eqnarray}
w_\xi(x)=\frac{\sqrt{x}}{\sqrt{x}+\xi \sqrt{1-x}}\ (0\le x \le 1).\label{PWF}
\end{eqnarray}
Figure.1 shows the behavior of $w_\xi (x)$ at $m=3$ and $\xi=\sqrt{m-1}$.
\begin{figure}[htb]
  \begin{center}
    \includegraphics[width=70mm,height=70mm]{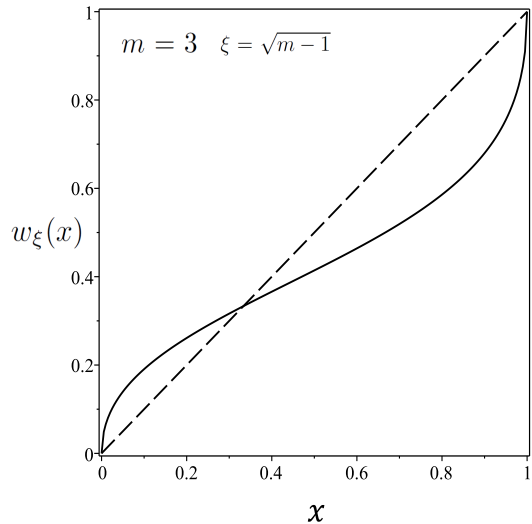}
  \end{center}
  \caption{The behavior of $w_\xi (x)$ at $m=3$ and $\xi=\sqrt{m-1}$ (solid line) that is similar to the probability weighting function proposed in the prosect theory.}
  \label{fig:pwf.eps}
\end{figure}
One can find that $w_\xi(x)$ realizes the character of \textit{probability weighting function}, which explains the experimental fact that small (statistical) probabilities are overestimated, and large probabilities are underestimated, subjectively. In the prospect theory, the probability weighting function is the important concept to explain the violation of
 independence axiom in VNM theory. Actually, from phenomenological discussions, various weighting functions has been proposed(\cite{PR,RW,TK}). Especially, we note the form of two-parameter weighting function, 
\[w_{\lambda, \delta}(x)=\frac{\delta x^{\lambda}}{\delta x^{\lambda}+(1-x)^{\lambda}},\] 
which was discussed in \cite{GW}. The parameters $ \lambda $ and $ \delta $ control the curvature and elevation of function, respectively. 
This phenomenological function with $ \lambda=1/2 $ and $ \delta=1/\xi $ is realized 
in the form of Eq.~(\ref{PWF}), which is derived from the state representation by density operator. \\
\par
In the discussions so far, we showed that the non-classical criterion $\varphi_D(\rho)$ had the two characteristics:
\begin{enumerate}
  \item $\varphi_D(\rho)$ includes the evaluation of risk.
  \item $\varphi_D(\rho)$ explains the property of probability weighting function.
\end{enumerate}
These are important points needed for the description of realistic selection behaviors.
Actually, in the below sections, $\varphi_D(\rho)$ is used to describe several famous decision-making phenomena, which are impossible to be explained in the framework of the standard expected utility theory.

\section{Description of Selection Behavior Under Ambiguity}
In this section, we consider selection behavior under \textit{unknown risk} whose probability distribution is not informed. 
Selection behavior under \textit{ambiguity} was first discussed by Ellsberg: He pointed out that many real decision makers prefer known
risks to unknown risks, and such a tendency called \textit{ambiguity aversion} cannot be explained in the standard expected utility theory.
To solve this paradox, several mathematical models have been proposed. 
However, Machina recently presented the example of selection behavior under ambiguity that cannot be explained even in the popular models of ambiguity aversion.
In Sec.3.1, we discuss the solution of Ellsberg paradox in our model, and in Sec.3.2, we show that \textit{our model also can solve the 
 Machina paradox} in the same manner.
 
\subsection{Solution of Ellsberg Paradox}
The Ellsberg Paradox is explained in the following sentence:\\
\textit{An urn contains $90$ balls. $30$ balls are red ($R$). The other $60$ balls are either white ($W$) or yellow ($Y$), but its ratio is unknown. Consider the four lotteries shown in Table.1, in each of which, if you draw one ball from the urn, you might get $\$100$.}
\begin{table}[tb]
 \caption{Example of Ellsberg Paradox.}
  \begin{center}
  \begin{tabular}{c|ccc}
            &    $30$balls  & \multicolumn{2}{c}{$60$balls} \\
            &  R  &  W  &  Y  \\
    \hline \hline
      $f_1$ & \$100   & 0   &   0 \\
    \hline
    $f_2$   & 0   &  \$100  &  0  \\
    \hline
     $f_3$  &  \$100  & 0   &  \$100  \\
    \hline
    $f_4$   &  0  &  \$100  &  \$100  \\
    \hline
  \end{tabular}
 \end{center}
\end{table}\\
Note, all of the lotteries consist of the event of getting \$100 and the one of getting nothing. The table.2 shows the probabilities that are assigned for the events,$\{(f_k,100),(f_k,0),k=1,2,3,4\}$.
\begin{table}[tb]
 \caption{Probabilities of events $(f_k,100)$ and $(f_k,0)$.}
 \begin{center}
  \begin{tabular}{|c|c|c|}
   \hline
            &  $(f_k,100)$  &  $(f_k,0)$    \\
    \hline  \hline    $f_1$  &  $1/3$  &  $2/3$    \\
    \hline 
     $f_2$  &  $2\alpha/3$  &   $(3-2\alpha)/3$    \\
    \hline
     $f_3$  &   $(3-2\alpha)/3$  &   $2\alpha/3$     \\
    \hline
     $f_4$  &  $2/3$  &  $1/3$    \\
    \hline
  \end{tabular}
 \end{center}
\end{table}
The parameter $\alpha$ ($0\le \alpha \le 1$) in the table specifies the unknown ratio of white and yellow balls, and it is included in the probabilities of $(f_k,100)$ and $(f_k,0)$ with $k=2,3$. 
One can see that the lots $f_2$ and $f_3$ have ambiguities on the probabilities of results.
Ellsberg predicted that many real decision makers prefer $f_1$ to $f_2$ and $f_4$ to $f_3$, that is, $(f_1\succ f_2) \wedge (f_4\succ f_3)$.
This tendency of ambiguity aversion is inconsistent with the standard manner according to the expected utility theory, in which, $f_1 \succ f_2$ if and only if $f_3\succ f_4$.   
The aim in this subsection is to simulate the selection behavior in the above situation by using the model in Sec.~2. Firstly, let us consider the comparison state given for the selection between $f_1$ and $f_2$. We use the form of non-classical state of Eq.~(\ref{QLC0});
\begin{equation}
\rho_{f_1-f_2}=|\Psi \rangle \langle \Psi|,\ \ket{\Psi}=\frac{1}{\sqrt{2}}(\ket{\psi_{f_1}}+\ket{\psi_{f_2}}),\label{EL12}
\end{equation}
where 
\begin{eqnarray}
&&\ket{\psi_{f_1}}=\sqrt{\frac{1}{3}}\ket{f_1}\otimes \ket{100}+\sqrt{\frac{2}{3}}\ket{f_1}\otimes \ket{0}=\sqrt{\frac{1}{3}}\ket{1}+\sqrt{\frac{2}{3}}\ket{2}, \nonumber \\
&&\ket{\psi_{f_2}}=\sqrt{\frac{2\alpha}{3}}\ket{f_2}\otimes \ket{100}+\sqrt{1-\frac{2\alpha}{3}}\ket{f_2}\otimes \ket{0}
=\sqrt{\frac{2\alpha}{3}}\ket{3}+\sqrt{1-\frac{2\alpha}{3}}\ket{4}.\nonumber \\ 
\end{eqnarray}
Here, $\{\ket{1},\ket{2},\ket{3},\ket{4}\}$ is a CONS on $\mathcal{H}=\mathbb{C}^4$, which corresponds to the set of events in the lots $f_1$ and $f_2$. 
Note, $\rho_{f_1-f_2}$ is rewritten as
\[\rho_{f_1-f_2}=\frac{1}{2}\sigma_{f_1}+\frac{1}{2}\sigma_{f_2}+\frac{1}{2}C_{f_1\leftrightarrow f_2},\]
with $\sigma_{f_k}=|\psi_{f_k}\rangle \langle \psi_{f_k}|$ and $C_{f_1\leftrightarrow f_2}=|\psi_{f_1}\rangle \langle \psi_{f_2}|+|\psi_{f_2}\rangle \langle \psi_{f_1}|$.\par
The operator $D$ in the evaluation function $\varphi_D$ is defined in the form of Eq.~(\ref{EF});
\begin{equation}
D=\sum_{i,j}^4 F(i,j)|i\rangle \langle j|,\nonumber
\end{equation}
where $F(i,j)^*=F(j,i)$. If $(i,j)=((f_k,100),(f_{k'},0))$, $F(i,j)=(u(\$ 100)-u(0))\rm{e}^{i\phi_{\it{ij}}}$ and 
if $(i,j)=((f_k,100),(f_{k'},100))$ or $((f_k,0),(f_{k'},0))$, $F(i,j)=0$.
In the above definitions, the criterion $\varphi_D(\rho_{f_1-f_2})$ is calculated as
\begin{eqnarray}
\varphi_D(\rho_{f_1-f_2})=\varphi_D \left(\frac{1}{2}\sigma_{f_1}\right)+\varphi_D\left(\frac{1}{2}\sigma_{f_2}\right)+\varphi_D \left(\frac{1}{2}C_{f_1\leftrightarrow f_2}\right),\nonumber
\end{eqnarray}
\begin{eqnarray}
&&\varphi_D \left(\frac{1}{2}\sigma_{f_1}\right)=\frac{\sqrt{2}}{3}\cos \Theta_{12} \delta u,\\
&&\varphi_D\left(\frac{1}{2}\sigma_{f_2}\right)=\frac{\sqrt{2\alpha(3-2\alpha)}}{3}\cos \Theta_{34}\delta u,\label{ELR}\\
&&\varphi_D \left(\frac{1}{2}C_{f_1\leftrightarrow f_2}\right)=\left(\frac{\sqrt{3-2\alpha}}{3}\cos \Theta_{14}+\frac{2\sqrt{\alpha}}{3}\cos \Theta_{23}\right)\delta u. 
\end{eqnarray}
Here $\delta u=u(\$100)-u(0)>0$. $\varphi_D \left(\frac{1}{2}C_{f_1\leftrightarrow f_2}\right)$ consists of $\frac{\sqrt{3-2\alpha}}{3}\cos \Theta_{14}\delta u$ 
and $\frac{2\sqrt{\alpha}}{3}\cos \Theta_{23}\delta u$, which are the terms of comparison between $(f_1,100)$ and $(f_2,0)$, and the one between $(f_1,0)$ and $(f_2,100)$, respectively.
The difference of utilities of $(f_1,100)$ and $(f_2,0)$ will contribute to increase the preference of $f_1$, and the difference of utilities of $(f_1,0)$ and $(f_2,100)$ will decrease it. From this perspective, $\cos \Theta_{14}\ge 0$ and  $\cos \Theta_{23}\le 0$ are assumed.
The terms of $\varphi_D \left(\frac{1}{2}\sigma_{f_1}\right)$ and $\varphi_D\left(\frac{1}{2}\sigma_{f_2}\right)$ represent the evaluations of risks in $f_1$ and $f_2$. For the decision maker who has the tendency to dislike (like) risks, the risk in $f_1$ will become a cause to decrease (increase) the preference of $f_1$, and the one in $f_2$ will be a cause to increase (decrease) it. Thus, the two degrees of evaluations of risks (DERs) given by $\cos \Theta_{12}$ and $\cos \Theta_{34}$ satisfy $\cos \Theta_{12} \cos \Theta_{34} \le 0$. 
Further, we have to note that $\varphi_D\left(\frac{1}{2}\sigma_{f_2}\right)$ has the unknown parameter $\alpha$, see Eq.~(\ref{ELR}),
that is, it is the evaluation of risk with ambiguity.
According to Ellsberg's prediction, generally, the unknown risk affects the selection behavior more strongly than the known risk.
This essence is represented in the condition $|\cos \Theta_{12}|<| \cos \Theta_{34}|$ in our model.  
To simplify the discussion, let us set the parameters $\set{\Theta_{ij}}$ as satisfying
\[\cos \Theta_{14}=1,\ \cos \Theta_{23}=-1,\ \cos \Theta_{12}=0,\ \cos \Theta_{34}=\lambda.\]
Then,
\begin{eqnarray}
\varphi_D(\rho_{f_1-f_2})=\left(\frac{\sqrt{3-2\alpha}}{3}-\frac{2\sqrt{\alpha}}{3}+\lambda \frac{\sqrt{2\alpha(3-2\alpha)}}{3}\right)\delta u. \label{EX12}
\end{eqnarray}
The parameter $\lambda$ is DER for the lot $f_2$ with ambiguity; if $\lambda>0$ ($<0$), the decision maker dislikes (likes) the ambiguity.
This criterion is used as $f_1\succ f_2$ if $\varphi_D(\rho_{f_1-f_2})>0$ or $f_2\succ f_1$ if $\varphi_D(\rho_{f_1-f_2})<0$. In the same manner, we can consider the criterion for the selection between $f_3$ and $f_4$ that is given by
\begin{eqnarray}
\varphi_D(\rho_{f_4-f_3})=\left(\frac{2\sqrt{\alpha}}{3}-\frac{\sqrt{3-2\alpha}}{3}+\lambda \frac{\sqrt{2\alpha(3-2\alpha)}}{3}\right)\delta u. \label{EX34}
\end{eqnarray}
Note, the comparison state $\rho_{f_4-f_3}$ is defined on the Hilbert space $\mathcal{H}'=\mathbb{C}^4$, which is different from $\mathcal{H}$. This criterion is used as $f_4\succ f_3$ if $\varphi_D(\rho_{f_4-f_3})>0$ or $f_3\succ f_4$ if $\varphi_D(\rho_{f_4-f_3})<0$.
The parameter $\lambda$ in the third term is DER for the lot $f_3$ with ambiguity.\par
The criterions shown in Eqs.~(\ref{EX12}) and (\ref{EX34}) are regarded as the functions of variables $\alpha$ and $\lambda$:
$\varphi_{D}(\rho_{f_1-f_2}):=S_{f_1-f_2}(\alpha, \lambda)$ and $\varphi_{D}(\rho_{f_4-f_3}):=S_{f_4-f_3}(\alpha, \lambda)$.
Thus, the decision maker's selection depends on the values of $(\alpha, \lambda)$. See Fig.~2 that shows $\alpha - \lambda$ phase diagram giving $(f_1\succ f_2)\wedge (f_3\succ f_4) $, $(f_1\succ f_2)\wedge (f_4\succ f_3) $, $(f_2\succ f_1)\wedge (f_4\succ f_3) $ or $(f_2\succ f_1)\wedge (f_3\succ f_4) $.
 
\begin{figure}[tb]
  \begin{center}
    \includegraphics[width=70mm,height=70mm]{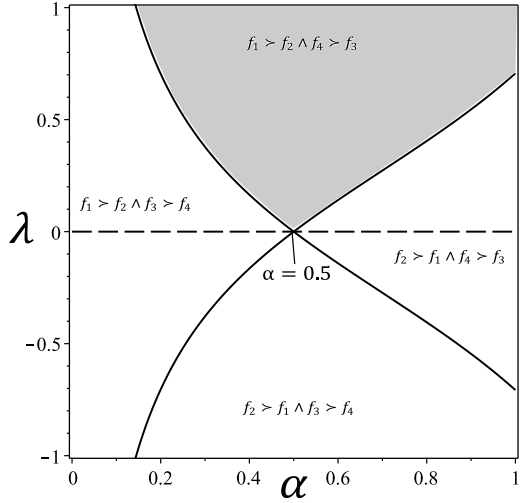}
  \end{center}
  \caption{$\alpha-\lambda$ phase diagram of selection in Ellseberg Paradox, in which, the parameter $\alpha$ specifies the unknown ratio and $\lambda$ is DER for the lots $f_2$ and $f_3$. Ellsberg's prediction ($(f_1\succ f_2) \wedge (f_4\succ f_3)$) is realized in the range of $\lambda>0$.}%
    \label{fig:1.eps}
\end{figure}
Let us see the line of $\lambda=0$ in this diagram: The preference $(f_1\succ f_2)\wedge (f_3\succ f_4)$ is obtained when $\alpha<1/2$, and $(f_2\succ f_1)\wedge (f_4\succ f_3)$ 
when $\alpha <1/2$. This is consistent with the result derived from the standard expected utility theory.
One can see that the preference $(f_1\succ f_2) \wedge (f_4\succ f_3)$, which was predicted by Ellsberg, appears in the range of $\lambda>0$. Then, the decision maker has the tendency to dislike the unknown risks in $f_2$ and $f_3$ and embeds this evaluation into his (her) selection behavior.
Especially, at $\alpha\approx 1/2$, the preference $(f_1\succ f_2) \wedge (f_4\succ f_3)$ is realized even if the positive $\lambda$ is very small.
In general, $\alpha$ and $\lambda$ are quantities that are fluctuated or determined subjectively by the decision maker. 
However, we predict that many people stay at $\alpha\approx 1/2$ unless some external information on the ratio of white and yellow balls is given.
In other words, we are forced to be aware of the existence of white and yellow balls fairly, in the presence of the fact that the ratio is unknown.  
From such a perspective, we conclude that Fig.~2 encourages Ellsberg's prediction. 

\subsection{Solution of the Machina Paradox}
 Machina presented the following situation as an example of selection under ambiguity~\cite{}:\\
\textit{An urn contains $101$ balls. 
The $50$ balls are either red ($R$) or black($B$). The $51$ balls are either white (W) or yellow (Y). 
These ratios are unknown. 
Consider the four lotteries shown in Table.~3, where you will get outcomes whose utilities are 303, 202 , 101 or 0, if you draw one ball from the urn.}
\begin{table}[tb]
 \caption{Example of the Machina Paradox.}
  \begin{center}
  \begin{tabular}{c|cccc}
            &   \multicolumn{2}{c}{$50$balls}  & \multicolumn{2}{c}{$51$balls} \\
            &  R  &  B  &  W  &  Y  \\
    \hline \hline
      $f_1$ & 202   & 202   &   101  & 101 \\
    \hline
    $f_2$   & 202   &  101  &  202  &101 \\
    \hline
     $f_3$  &  303  & 202  &  101 & 0\\
    \hline
    $f_4$   &  303  &  101  &  202 & 0\\
    \hline
  \end{tabular}
 \end{center}
\end{table}

\begin{table}[tb]
 \caption{Probabilities of events $(f_k,3a)$, $(f_k,2a)$, $(f_k,a)$, $(f_k,0)$ }
 \begin{center}
  \begin{tabular}{|c|c|c|}
   \hline
            &  $(f_k,2a)$  &   $(f_k,a)$    \\
    \hline  \hline    
    $f_1$  &  $p$  &  $q$    \\
    \hline 
     $f_2$  &  $p\alpha +q\beta$  &   $1-(p\alpha +q\beta)$    \\
    \hline
  \end{tabular}
 \end{center}
 
 \begin{center}
  \begin{tabular}{|c|c|c|c|c|}
   \hline
            &  $(f_k,3a)$  &  $(f_k,2a)$  & $(f_k,a)$ & $(f_k,0)$ \\
    \hline  \hline    
    $f_3$  &  $\ \ \ p\alpha \ \ \ $  &  $p(1-\alpha)$  & $q\beta$ & $q(1-\beta)$ \\
    \hline 
     $f_4$  &  $\ \ \ p\alpha  \ \ \ $  & $q\beta$   & $p(1-\alpha)$ & $q(1-\beta)$  \\
    \hline
  \end{tabular}
 \end{center}
 
\end{table}
The lots $f_1$ and $f_2$ consist of $(f_k,2a)$ and $(f_k,a)$ with $a=101$. The lots $f_3$ and $f_4$ consist of $(f_k,3a)$, $(f_k,2a)$, $(f_k,a)$ and $(f_k,0)$.
The table.~4 shows the probabilities assigned for these events, where $p=50/101$, $q=1-p=51/101$ and the parameters $\alpha,\beta$ ($0\le \alpha,\beta \le 1$) specify the unknown ratio of $R$ and $B$ and the one of $W$ and $Y$.
One can see that only the lot $f_1$ is unambiguous, and other lots have ambiguity on the probabilities of events.
Machina pointed out that this example falsifies the popular models of ambiguity aversion including Choquet expected utility(\cite{SC}), maxmin expected utility(\cite{GS}), variational preferences(\cite{MA}), $\alpha$-maxmin(\cite{GH}) and the smooth model of ambiguity 
aversion(\cite{KL}).  According to these models, $f_3\succ f_4$ if and only if $f_1\succ f_2$. Since $f_1$ is unambiguous, the preference $f_1\succ f_2$ can be interpreted as the result of the ambiguity aversion. However, both of $f_3$ and $f_4$ include ambiguities. Note, the lotteries $f_2$ and $f_4$ have a slight advantage due to the $51$th ball that may yield $202$. In the choice between $f_1$ and $f_2$, there is a tradeoff between the advantage offered by $f_2$ and the unambiguity offered by $f_1$. 
On the other hand, such trade-off seems to be less in the selection of $f_3$ or $f_4$. From this perspective, $(f_1\succ f_2) \wedge (f_4\succ f_3)$ can not be denied as a realistic selection.\par
With using the manner discussed in Sec.2.1, we analyze the selection behavior in this situation.
Criterion for the selection between $f_1$ and $f_2$ is designed as 
\begin{eqnarray}
\varphi_D(\rho_{f_1-f_2})=\varphi_D\left(\frac{1}{2}\sigma_{f_1}\right)+\varphi_D\left(\frac{1}{2}\sigma_{f_2}\right)+\varphi_D\left(\frac{1}{2}C_{f_1\leftrightarrow f_2}\right),
\end{eqnarray}
where
\begin{eqnarray}
&&\varphi_D\left(\frac{1}{2}\sigma_{f_1}\right)=0,\label{12}\\
&&\varphi_D\left(\frac{1}{2}\sigma_{f_2}\right)=\sqrt{(p\alpha+q\beta)(1-(p\alpha +q\beta))}\lambda a,\label{34}\\
&&\varphi_D\left(\frac{1}{2}C_{f_1\leftrightarrow f_2}\right)=\left(\sqrt{p(1-(p\alpha +q\beta))}-\sqrt{q(p\alpha +q\beta)}\right)a.\label{com12}
\end{eqnarray}
We assume, the DER for the lot $f_1$ with the known risk is zero, see Eq.~(\ref{12}), and the DER for the lot $f_2$ with the unknown risk is $\lambda$, see Eq.~(\ref{34}).
All of the comparisons of utilities are embedded in the term of Eq.~(\ref{com12}).
We define this criterion as a function of $\alpha$, $\beta$ and $\lambda$; $\varphi_D(\rho_{f_1-f_2}):=S_{f_1-f_2}(\alpha,\beta,\lambda)$.
If $S_{f_1-f_2}(\alpha,\beta,\lambda)>0\ (S_{f_1-f_2}(\alpha,\beta,\lambda)<0)$, $f_1\succ f_2$ ($f_2\succ f_1$).
Next, the criterion for the selection between $f_3$ and $f_4$ is designed as
\begin{eqnarray}
\varphi_D(\rho_{f_4-f_3})=\varphi_D\left(\frac{1}{2}\sigma_{f_4}\right)+\varphi_D\left(\frac{1}{2}\sigma_{f_3}\right)+\varphi_D\left(\frac{1}{2}C_{f_4\leftrightarrow f_3}\right),
\end{eqnarray}
where
\begin{eqnarray}
&&\varphi_D\left(\frac{1}{2}\sigma_{f_4}\right)=-2\left(p\sqrt{\alpha(1-\alpha)}+q\sqrt{\beta(1-\beta)}\right)\lambda a,\label{12-34}\\
&&\varphi_D\left(\frac{1}{2}\sigma_{f_3}\right)=\left(p\sqrt{\alpha(1-\alpha)}+q\sqrt{\beta(1-\beta)}\right)\lambda a,\label{57-68}\\
&&\varphi_D\left(\frac{1}{2}C_{f_4\leftrightarrow f_3}\right)=\left(\sqrt{pq\alpha \beta}+q\sqrt{\beta (1-\beta)}+q\beta \right)a \nonumber \\ && \ \ \ -\left(\sqrt{pq(1-\alpha)(1-\beta)}+p\sqrt{\alpha (1-\alpha)}+p(1-\alpha)\right)a.\label{com43}
\end{eqnarray}
Note, the DERs for $f_3$ and $f_4$ with the unknown risks are also specified by the parameter $\lambda$, see Eqs.~(\ref{12-34}) and (\ref{57-68}).
If $\varphi_D(\rho_{f_4-f_3}):=S_{f_4-f_3}(\alpha,\beta,\lambda)>0$ ($S_{f_4-f_3}(\alpha,\beta,\lambda)<0$), $f_4\succ f_3$ ($f_3\succ f_4$). 
\begin{figure}[tb]
  \begin{center}
    \includegraphics[width=70mm,height=70mm]{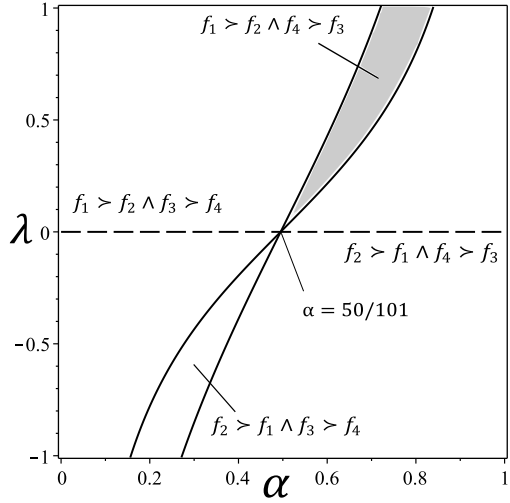}
  \end{center}
  \caption{$\alpha-\lambda$ Phase Diagram of Selection in the Machina Paradox}
  \label{fig:5.eps}
\end{figure}
Figure.~3 shows $\alpha - \lambda$ phase diagram in the case of $\alpha=\beta$.
Let us see the line of $\lambda=0$ in this figure. If $\alpha>50/101$, the selection $(f_2\succ f_1)\wedge (f_4\succ f_3)$ is realized.
As mentioned in the previous section, the decision maker will stay at $\alpha=\beta\approx 1/2$, unless some information on the unknown ratios is given. 
Since $1/2>50/101$, the one selects $f_2$ and $f_4$; in a sense, the difference between $1/2$ and $50/101$ represents the slight advantage of $f_2$ and $f_4$, which comes from the belief that the 51th ball might be $W$.
When $\lambda$ becomes large, $(f_1\succ f_2)\wedge (f_4\succ f_3)$ and $(f_1\succ f_2)\wedge (f_3\succ f_4)$ are appearing. 
As mentioned previously, the positive $\lambda$ implies the tendency of ambiguity aversion which decreases the values of $f_2$, $f_3$ and $f_4$ with ambiguity.
As seen in the diagram of Fig.~3, the value of $\lambda$ needed to invert $f_2\succ f_1$ to $f_1\succ f_2$ is not so large, but 
$\lambda$ to invert $f_4\succ f_3$ to $f_3 \succ f_4$ has the higher value. 
The suppression of inversion from $f_4\succ f_3$ to $f_3 \succ f_4$ is due to the competition between the evaluations of unknown risks offered by the both lots, see 
Eqs.~(\ref{12-34}) and (\ref{57-68}). As a result, our model explains the existence of selection 
$(f_1\succ f_2)\wedge (f_4\succ f_3)$, which has not been explained in the popular models of ambiguity aversion.

\section{Decision of Cash Equivalent}
In this section, we consider \textit{cash equivalent}(CE) for lot with risk
that means amount of money whose utility is indifferent with the utility of receiving lot:
\[u_{CE}=u_{LOT}.\]
This relation can be interpreted as a result of comparison between the two alternatives, the selection of definite cash and the one of lot. Thus, the decision of CE can be discussed in the framework of our comparison model. 
For example, let us consider a comparison between a cash $x$ and a lot whose outcome is $y(>0)$ with probability $p$ or nothing with $q=1-p$.
Then, the comparison state is given as
\begin{eqnarray}
\rho=|\Psi \rangle \langle \Psi |,\ \ket{\Psi}=\frac{1}{\sqrt{2}}\ket{\psi_{lot}}+\frac{1}{\sqrt{2}}\ket{\psi_{cash}},
\end{eqnarray}
where
\begin{eqnarray}
&&\ket{\psi_{lot}}=\sqrt{p}\ket{L}\otimes \ket{y}+\sqrt{q}\ket{L}\otimes \ket{0}=\sqrt{p}\ket{1}+\sqrt{q}\ket{2},\nonumber\\
&&\ket{\psi_{cash}}=\ket{C}\otimes \ket{x}=\ket{3}.
\end{eqnarray}
The comparison state $\rho$ is defined in $\mathcal{H}=\mathbb{C}^3$. With the same manner in the previous sections, we design the following criterion. 
\begin{eqnarray}
\varphi_{D}(\rho)=\sqrt{p}(u_y-u_x)-\sqrt{q}u_x-\lambda \sqrt{pq}u_y.
\end{eqnarray}
$u_x$ and $u_y$ are the utilities of the outcomes $x$ and $y$, and $u_0=0$ is assumed.
The parameter $\lambda$ is the DER for the lot.
If $\varphi_D(\rho)=0$, the lot and $x$ are indifferent for the decision maker.
We define the cash $x$ in this case as the CE of the lot;
\begin{eqnarray}
u_{CE}=u_x=\frac{\sqrt{p}(1-\lambda \sqrt{q})u_y}{\sqrt{p}+\sqrt{q}}.\label{CE}
\end{eqnarray}
\begin{figure}[tb]
  \begin{center}
    \includegraphics[width=75mm,height=75mm]{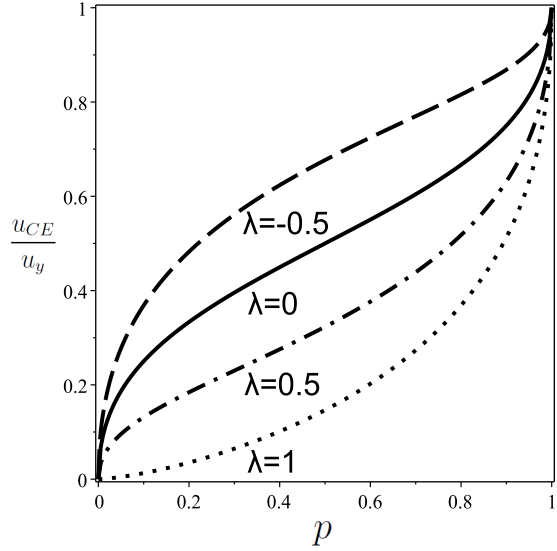}
  \end{center}
  \caption{Behavior of cash equivalent of lot.}
  \label{fig:CE.eps}
\end{figure}
Figure.~4 shows the behavior of $u_{CE}/u_y$ for the parameters of $p$ and $\lambda$ ($0< p, \lambda < 1$).
One can see that for any $p$, $u_{CE}$ is smaller than $u_y$, which is the utility of the highest outcome in lot, and its value is decreased as $\lambda$ becomes large. Note, the decision maker at positive $\lambda$ dislikes the risk of lot.\par
The CE defined in Eq.~(\ref{CE}) will be closely related to \textit{willingness to accept} (WTA) and \textit{willingness to pay} (WTP) for the lot, because, each of them is also defined as cash equivalent to the lot. 
WTA and WTP are conceptually distinguished in the difference situations that require the determination of CE. 
WTA is the minimum cash which the decision maker accepts to sell the lot.
WTP is the maximum cash which the decision maker pays for getting the lot.
In the determination of WTA, the decision maker is \textit{endowed} with the lot.
On the other hand, in the case of WTP, the decision maker does not get the lot yet.
In our model, this apparent difference of position is reflected on the value of $\lambda$ in Eq.~(\ref{CE}).
The seller abandons the right of drawing the lot, therefore, he/she might not fear the underlying risk in the lot.
The buyer will be aware of the risk more strongly, because he/she might draw the lot.
From such a perspective, it is expected that  seller's $\lambda$ will be smaller than the one the buyer has.
As seen in Fig.~4, if $\lambda<\lambda'$,
\[CE(\lambda)>CE(\lambda')\]
is always satisfied.
Actually, WTA $>$ WTP is the fact that has been confirmed in various experimental tests.
Also, this discussion is closely related to \textit{preference reversal phenomena} (\cite{TVT}). It is the following decision making:
There are two lots, one of which, say $ A $, has a high chance of winning a small prize, and another, say $ B $, has a lower chance of winning a larger. According to experimental tests, many people prefer $ A $ to $ B $ ($ A\succ B $) but, they estimate WTA of $ B $ larger than the one of $ A $. This fact seems to show the violation of transitivity axiom: In the situation of selling the lots, there exist a cash $ X $ satisfying the preference as $ B\succ X \succ A $, in spite of $ A \succ B $ in the simple comparison 
of the two lots. The essence of preference reversal is almost same with the disparity between WTA and WTP: In the situation of selling, 
they will estimate WTA with using $ \lambda $, which is smaller than $ \lambda' $ expected to be used in the simple 
comparison between $ A $ and $ B $.\par
Here, we mention the popular interpretations on the disparity between WTA and WTP: 
In the standard economic theory, the \textit{income effect} has been thought to be a cause of the disparity.\footnote{We remark the real situation is even more complicated:
in standard economics there are typically considered two effects - the income effect and substitution effect.} WTP is generally 
constrained by the present income, but WTA is not. However, the existence of disparity is found in many experimental tests, in which, 
income effects are expected to be small(\cite{CA,KA2}).  On the other hand, in the prospect theory, the disparity is explained from \textit{the difference of reference point in the value function}.
The value function evaluates gain or loss of outcome from reference point that represents a decision maker's situation. The decision maker endowed with the lot and the one not endowed are posited at different reference points. From these points, the compensation of loss of lot (WTA) and the cost of gain of lot (WTP) are estimated respectively. Generally, in the value function, the degree of depression of utility by loss becomes larger than the one of enhancement of utility by gain. Therefore, WTA$>$WTP is realized.\par
In our model, the existence of the parameter $\lambda$ (DER) explains the disparity between WTA and WTP, and it is a psychological factor, 
which fluctuates  depending on the decision maker's situation. This point is consisting with the concept of value function in the prospect theory, where the reference 
point fluctuates. Furthermore, the evaluation of loss or gain in the terms of prospect theory is clearly represented as the function of $\lambda$.

\section{Conclusion}
The decision maker simulated in our model performs various comparisons of objects to make a criterion of selection. As discussed in Sec.~2, all of the comparisons are embedded into the comparison 
state $ \rho $, and the  criterion is defined by the function $ \varphi(\rho) $. This quantum-like model has potential to explain realistic selection behaviors. Actually, it clearly explains the behavior of probability weighting function, see Fig.~1. In the prospect 
theory, the probability weighting function is assumed in a phenomenological sense and used to explain the violation of independence axiom in VNM theory. Further, our model describes comparisons, which are realistic operations but have been ignored in the expected utility theory. The evaluation
 of such comparison is reflected in the criterion $ \varphi(\rho) $, as the degree of evaluation of risk (DER). As discussed in Sec.~3, DER plays the important role to solve Ellsberg paradox and Machina paradox, see the diagrams of Figs.~2 and 3 where the parameters $ \alpha$ and $ \lambda $ specify the unknown ratio and DER respectively. Here, we predicted that many people would pay fair attentions to the unknown probabilities and dislike the risk, that is, $ \alpha\approx 1/2 $ and $ \lambda >0 $. The decision maker at this position has the tendency of ambiguity aversion and can realize the choice that 
Machina pointed out. Our solution is crucial, because the Machina paradox is difficult to be solved in the popular models of 
ambiguity aversion. (Our solution might be closely related to the discussion in the paper of \cite{Bai}, in which, models of decision making 
are discussed to account for the Machina paradox.) In Sec.~4, the disparity between WTA(willingness to accept) and WTP(willingness to pay) is explained. 
Here, we assumed that DER is the sensitive quantity varying depending on the situation of ``accept" or ``pay", see Fig.~4. 
This property is closely related to the preference reversal phenomenon that shows anomaly violating the transitivity axiom in VNM theory. More generally, DER might specify personalities including habits, experiences and cognitive abilities. This point will be very important from the view of experimental psychology.\par
Lastly, let us point out that this general modeling is applicable to various decision making phenomena. 
For example, it can address more complicated case such that a large number of alternatives exist. 
Naturally, a real decision maker cannot pay attention to all alternatives, and such a situation will be reflected on the form of comparison state. 
The problem on limited attention has been often discussed in the economic literatures, see the papers of \cite{MaMa} and \cite{Masa}.





\end{document}